# 3D Characterisation of the Fe-rich intermetallic phases in Al-5%Cu alloys by synchrotron X-ray microtomography and skeletonisation

Y. Zhao [a, b], W. Du [b], B. Koe [b, c], T. Connolley [c], S. Irvine [c], P. K. Allan [c], C. M. Schlepütz [d], W. Zhang [a], F. Wang [e], D.G. Eskin [e] and J. Mi *[b]

[a] National Engineering Research Centre of Near-Net-Shape Forming for Metallic Materials, South China University of Technology, Guangzhou, 510641, China

[b] School of Engineering & Computer Science, University of Hull, East Yorkshire, HU6 7RX, UK

[c] Diamond Light Source, Harwell Science and Innovation Campus, Didcot, Oxfordshire, OX11 0DE, UK

[d] Swiss Light Source, Paul Scherrer Institute, 5232 Villigen, Switzerland

[e] BCAST, Brunel University London, Uxbridge, UB8 3PH, UK

Corresponding author: J.Mi@hull.ac.uk (J. Mi)

**Abstract**

Synchrotron X-ray microtomography and skeletonisation method were used to study the true 3D network structures and morphologies of the Fe-rich intermetallic phases in Al-5.0%Cu-0.6%Mn alloys with 0.5% and 1.0% Fe. It was found that, the Fe-phases in the 1.0%Fe alloy have node lengths of 5-25μm; while those in the 0.5%Fe alloy are of 3-17 μm. The Fe-phases in the 1.0%Fe alloy also developed sharper mean curvature with wider distribution than those in the 0.5%Fe alloy. Combining SEM studies of the deeply-etched samples, the true 3D structures of 4 different type Fe-phases in both alloys are also revealed and demonstrated.

Aluminium (Al) alloys are widely used in the transportation, building and packaging industry because of their lightweight, high specific strength, high corrosion resistance, and excellent recyclability [1]. In modern vehicles, Al alloys are playing increasingly important roles in reducing the weight of vehicles and, hence, fuel consumption and $CO_2$ emissions in transportation [2]. Approximately 90% of the Al alloys used in land vehicles are from recycled sources for cost reduction and sustainability [3]. In Al alloys, especially recycled Al alloys where Fe concentration is often higher than 0.5% (weight percentage), Fe is the most common impurity element, and it can be easily picked up in sorting and remelting processes during Al recycling [4]. Normally, when the Fe in an Al alloy is >0.05% [4], brittle Fe-rich intermetallic phases (named Fe-phases hereafter) form and their size, morphology and distribution have profound effects on the castability and mechanical properties of the final parts. In most cases, these Fe-phases, especially when the needle-like or plate-like phases, such as β-$Al_7Cu_2Fe$ phase, are detrimental to the alloys [5]. In some alloy systems, neutralisation elements, *e.g.*, Mn and Si, can be used to alter the morphology of the Fe-phases to a less harmful type [6- 8].

Quantitatively understanding of the size, morphology and distribution of the Fe-phases are of paramount importance in the physical metallurgy of recycling Al alloys, and in manufacture high-quality components for the transportation industry. In the past, majority of the research on Fe-phases was conducted using 2-dimensional (2D) imaging methods [6-8], *i.e.* optical and/or electron microscopy, which gives very limited information about 3-dimensional (3D) structures/morphologies, and the spatial interconnections and correlations between the different phases. Recently, a number of investigations [9, 10] have been made by using scanning electron microscopy (SEM) and focused ion-beam (FIB) tomography to reveal the connected and branched 3D network structure in Chinese script type α-Fe-phases in Al-Si alloys (the typical composition is $Al_{14}Fe_{2.8}Si_2$). 3D morphology of Fe-phases has also been characterised using serial sectioning plus optical [11] or electron microscopy [12]. However, FIB is normally used for sectioning sub-micrometre features [12], not for those of length scale in many hundreds, even thousands of micrometres, such as the Fe-phases in present study, and serial sectioning is often very time-consuming. Recently, synchrotron X-ray tomography has been used to study the 3D microstructures of a wide range of multiphase alloys [13-16]. For example, the nucleation and growth of the Fe-phases in 3D in Al-Si alloys were reported in [4, 17-20]; and the snapshots of the 3D Fe-phases in Al-Cu alloys were given by Gutiérrez, *et al* [21]. However, Gutiérrez, *et al* did not segment the individual Fe-phases [21]. Hence, the detailed 3D structures of the Fe-phases, and their spatial interconnection with other phases such as $Al_2Cu$ have not been revealed. Normally, 4 different types Fe-phases exist in the Al-Cu alloys with Fe concentration of 0.5 - 1.0% [22, 23]. They are plate-shaped phases β-$Al_7Cu_2Fe$ and $Al_3(FeMn)$; and Chinese script-type phases, α-$Al_{15}(FeMn)_3Cu_2$, and $Al_6(FeMn)$. These Fe-phases are very different to the plate-shaped Fe-phases found in the Al-Si alloys [4, 17-20]. So far, no reports have been found that describe the true 3D structures of the 4 typical Fe-phases present in the Al-Cu alloys [22, 23] and their spatial interconnections and correlations.

In this paper, we used synchrotron X-ray microtomography and skeletonisation method to study the 3D network structures and morphologies of the Fe-phases and the associated $Al_2Cu$ phases in two alloys: Al-5%Cu-0.6%Mn with 0.5% and 1.0% Fe (named 0.5Fe alloy and 1.0Fe alloy, respectively, hereafter). Higher Fe content was deliberately added into the two alloys to mimic those often found in the recycled Al alloys. The complex 3D network structures of the Fe-phases and the $Al_2Cu$ phases, their mean curvature distributions and the inter-dependence between the Fe-phases and the $Al_2Cu$ phases obtained by skeletonisation analyses were reported for the first time. Furthermore, the true 3D morphologies of the 4 different types Fe-phases in Al-5%Cu alloys are also revealed, providing more quantitative 3D information for understanding the structures of the Fe-phases.

The 0.5Fe and 1.0Fe alloys were made by using pure Al ingot (99.9%), Al-20% Cu, Al-10%Mn and Al-10%Fe master alloy (provided by Sichuan Lande Industry Co., Ltd., China) with the correct charge weight. The feedstock materials were held inside a clay-graphite crucible and heated and melted at 780 °C in an electric furnace. Then, the alloy melt was degassed at 750 °C by submersing 20g of degas agent ($TiO_2$ powder mixed with 0.5% $C_2Cl_6$ powder wrapped by an Al foil) into the melt for 10 min. The melt was then cooled to 710 °C, and poured into a steel permanent mould (Ø 65 mm × 70 mm) preheated to 200 °C to form an ingot. Cylindrical samples (~Ø10 mm × 20 mm) were cut from the edge of the ingot and then machined into Ø 2 mm × 5 mm for tomography scans. The solidification time at the location where the samples were taken was ~ 42.5 s with an average cooling rate of ~ 2.5 K/s [24]. Routine 2D microstructure characterisation was made using a FEI Quanta 200 Field Emission Gun scanning electron microscope (SEM) equipped with an energy-dispersive X-ray analyzer. For the SEM samples, 10% NaOH aqueous solution was used to dissolve the Al matrix (20 min) in order to expose more of the Fe-phases embedded inside the Al matrix. Synchrotron X-ray tomography experiments were performed at the TOMCAT beamline X02DA of the Swiss Light Source (SLS), Paul Scherrer Institute, Switzerland. The experimental parameters used are given in Table 1 [25]. A white beam from a superconducting bending magnet source was used with a 400 µm Al filter to remove the low energy tail of the incident beam and to reduce heat load on the sample and detector. The imaging system consists of a 100 µm LuAG: Ce scintillator (Crytur) coupled to a white-beam compatible microscope with a 6.8 × magnification (Optique Peter). For each scan, 2000 projections were acquired over 180° of sample rotation.

Tomographic reconstructions were performed on the TOMCAT cluster [26] using the GridRec algorithm [27] coupled with the Parzen filter [28].

Fig. 1a shows the typical cross-sectional slice obtained from the synchrotron X-ray tomoscan, and an area of interest was extracted and showed in Fig. 1b. The corresponding SEM images and the deeply-etched Fe-phases are shown in Fig. 1c and d respectively. Both X-ray images and SEM images show that the pore is in dark, the Al matrix is in dark grey, the Fe-phases are in light grey, and the $Al_2Cu$ phases are in white colour. Open source image processing software, Image J [30] was used to adjust the contrast between the different phases. Then, the 3D bilateral filter was applied to the tomography datasets to increase the contrast and reduce noise. Finally, the pores, Al dendrites, Fe-phases, $Al_2Cu$ phases were manually segmented by using different global threshold values (Pore: 0 ~ 10688, α-Al: 10689 ~ 26438, Fe-phases: 26439 ~ 38814; $Al_2Cu$: 38815 ~ 65535). 3D segmentation and feature rendering were performed using Avizo Lite v9.0.1 (VSG, France) and Viper, the University of Hull's High Performance Computer (HPC) cluster. The image substacks obtained from Image J were reconstructed in 16-bit format. Normally, a region of interest sub-volume $100^3$ voxels in $500^3$ voxels with a voxel size of (1.62 micron)$^3$ was chosen for further analyses. The segmented were individually separated and labelled based on the connected regions. The 3D morphology of different phases were then smoothed by the value 1.5. The quantitative analysis (mean curvature, skeletonisation) for different phases were performed in the volume of $100^3$ voxels. However, from the X-ray absorption contrast only, it is not possible to distinguish and segment the 4 different types of Fe phases in the two alloys. Hence, the segmented Fe-phases from the X-ray tomoscans contain all 4-type Fe-phases. By comparing the 3D morphology of the Fe-phases partially revealed by the deeply-etched samples with those showed in the X-ray tomography, we are able to identify the 4 different type Fe-phases as discussed later in the paper.

Fig. 2 shows the 3D colour rendering of the Fe phases, Al$_2$Cu phases and α-Al matrix and their mean curvature distributions for the 0.5Fe and 1.0Fe alloys, respectively. The mean curvature $H$ [31] is defined as:

$$H = 0.5 * \left(\frac{1}{R_1} + \frac{1}{R_2}\right) \quad (1)$$

where $R_1$ and $R_2$ are the two principal radii of curves respectively. Local curvature is an important geometrical parameter for the interface between two phases (dendrites or intermetallics) formed during the solidification processes, influencing the diffusion of solutes and therefore the final morphology of the phases.

Fig. 2a and b show the complex network and intricate morphology of the interconnected Fe-phases and Al$_2$Cu phases. Red shows the Fe-phases, green for the Al$_2$Cu phases, blue for the Al matrix. These phases conglomerate together in the α-Al inter-dendritic region in the chosen volume of 162μm × 162μm ×162μm. Fig. 2c and d show the 3D network of the Fe-phases with their mean curvatures for the 0.5Fe and 1.0Fe alloy, respectively. The Fe-phases form a spatially interconnected complex 3D network. The distributions of their mean curvatures follow the Gaussian distribution (Fig. 2g). The distribution peak position (μ) of the Fe-phases increase from 0.64 to 1.10 and the standard deviation (σ) increases from 1.33 to 2.71 as the Fe increases from 0.5% to 1.0%. The Fe-phases in the 1.0Fe alloy have more positive and negative mean curvatures. Similarly, Fig. 2e and f show that the structures of the Al$_2$Cu phases are also interconnected 3D network. Fig. 2h shows that the mean curvatures of the Al$_2$Cu phases also follow Gaussian distribution. The distribution peak position (μ) increases from 0.34 to 0.75 as the Fe increases from 0.5% to 1.0%, and the standard deviation (σ) increases from 0.71 to 2.52, indicating that Al$_2$Cu phases of more positive mean curvatures also exist in the 1.0Fe alloy. *Much richer and clearer 3D information in different view angles for the Fe-phases and Al$_2$Cu phases are illustrated in the four companying videos*. The increase of the mean curvature is mostly due to the different solidification reactions in the 0.5Fe alloy and 1.0Fe alloy. In the 1.0Fe alloy, the solidification reactions in the range of 649-653 ºC are: L → α-Al; L → α-Al + $Al_3$(FeMn); L → α-Al + $Al_6$(FeMn) [22]; then at 542 ºC: L → α-Al + β-$Al_7Cu_2Fe$ + $Al_3$(FeMn) + $Al_6$(FeMn) + $Al_2Cu$ [22, 32]. The $Al_3$(FeMn) and $Al_6$(FeMn) phases are more rod-like phases with sharp edges as discussed later, resulting in a high proportion of positive mean curvature. While in 0.5Fe alloy, the solidification reactions in the range of 589-597 ºC are: L + $Al_6$(FeMn) → α-$Al_{15}$(FeMn)$_3Cu_2$; L + α-$Al_{15}$(FeMn)$_3Cu_2$ → β-$Al_7Cu_2Fe$ [22], and then in 542-537 ºC: L → α-Al + α-$Al_{15}$(FeMn)$_3Cu_2$ + β-$Al_7Cu_2Fe$ + $Al_2Cu$ [23, 32]. α-$Al_{15}$(FeMn)$_3Cu_2$ have typical Chinese script morphology, leading to a relatively low proportion of positive curvatures. In both alloys, the $Al_2Cu$ phases formed through the multiple-phase eutectic reaction and in close contact with the Fe-phases and the α-Al formed prior to the eutectic reaction. Hence, the $Al_2Cu$ phases somehow "inherit" the characteristics of the Fe-phases, i.e. more "flat" (near zero mean curvatures) Fe-phases resulted in more "flat" $Al_2Cu$ phases as in the 0.5Fe alloy case; while more "sharp" (higher positive or higher negative curvatures) Fe-phases led to more "sharp" $Al_2Cu$ phases as in the 1.0Fe alloy case.

We used the skeletonisation function available in Avizo® to peel off the 3D network of the Fe and the $Al_2Cu$ phases down to a skeleton (1-voxel thickness) with connecting nodes. The length of the curve between each node, the original thickness of the curve before the thinning process, and the number of the connecting nodes can be calculated, and therefore the 3D characteristics of skeleton (*the 3D nature of the phase branches, it should note that, for the phases of relatively flat shape, the skeletonisation can only pick up their edges as demonstrated more clearly in the companying videos*) can be quantified [31]. Fig. 3a and b show the skeletons of the Fe-phases in the 0.5Fe and 1.0Fe alloys, respectively. The Fe-phases in the 1.0Fe alloy have more branches. The node length distributions (Fig. 3e) show that the length of Fe-phases in the 0.5Fe alloy is shorter (in the range of 3-17 µm) than those (5-25µm) in the 1.0Fe alloy, indicating that Fe-phases of the 0.5Fe alloy are much compacted than those in the 1.0Fe alloy. This is consistent with the findings reported in [6] that the Fe-phases become less compact with increasing Fe concentration. Fig. 3c and d show the skeletons of the $Al_2Cu$ phases in the 0.5Fe and 1.0Fe alloys, and the distributions of the node lengths are shown in Fig. 3f. The skeleton structures of $Al_2Cu$ become more complex with the increase of Fe. The node length of $Al_2Cu$ in the 0.5Fe alloy is in the range of 5-20 μm, while the length for 1.0Fe alloys is in the range of 3-15 μm, indicating that the node length of $Al_2Cu$ in the 1.0Fe alloy is relatively shorter than those in the 0.5Fe alloy. This is related to the increase of the volume fraction of Fe-phases with increasing the Fe content [22], which occupy more room and consumed more Cu in the residual liquid before the melt approaches the eutectic reaction at ~546 °C, and then the $Al_2Cu$ phases form through the eutectic reaction. Hence the growths of $Al_2Cu$ phases are restricted within the space enclosed by the Fe-phases and the Al dendrites, resulted in an overall shorter length.

SEM-EDX analysis shows that composition of the 4 different type Fe-phases are: α-$Al_{15}(FeMn)_3Cu_2$ (Al: 77.43±2.62%, Fe: 12.09±2.73%, Mn: 3.74±0.87%, Cu: 6.73±2.32%, at.%); β-$Al_7Cu_2Fe$ (Al: 72.96±2.44%, Cu: 20.92±0.51%, Mn: 1.87±0.49%, Fe: 4.24±0.10%), $Al_3(FeMn)$ (Al: 81.58±0.87%, Cu: 5.52±1.40%, Mn: 2.32±0.38%, Fe: 10.56±1.01%) and $Al_6(FeMn)$ (Al: 83.67±0.86%, Cu: 3.30±0.47%, Mn: 3.17±0.01%, Fe: 9.86±0.09 %). Fig. 4 shows the volume rendering of 3D morphologies of the 4 type Fe-phases extracted from the tomographic images by carefully comparing with the deeply-etched phases obtained from the SEM observation. Fig. 4a shows the morphology of the $Al_3(FeMn)$ phases. In 3D, it is more or less like a long rod-like structure with a few small branches. While in 2D characterisation, they were often recognized as "needle-shaped phases" in the longitudinal direction. It is formed during the reaction: L → α-Al + $Al_3(FeMn)$. Fig. 4b shows that the $Al_6(MnFe)$ phase develops more branches, and become more complex in 3D. While the deeply-etched image further confirms such features. Fig. 4c reveals that the α-Fe is 3D complex network and apparently a typical 2D cross-sectional through the network could render a typical "Chinese script" in a 2D view field. This complex morphology is because $Al_6(FeMn)$ transforms to α-Fe through the peritectic reaction: $L + Al_6(FeMn) \rightarrow \alpha\text{-}Al + \alpha\text{-}Al_{15}(FeMn)_3Cu_2$ at 589-597 °C [22]. Thus, $Al_6(FeMn)$ is the nucleation site for the α-Fe phase. The α-Fe morphology presented in this study is significantly different from previous studies [6, 21, 22] reported in Al-Cu alloys, perhaps because those previous studies only observed a small volume of the deeply-etched samples [22] or FIB cut samples [10]. The $\beta\text{-}Al_7Cu_2Fe$ phase in both deeply-etched SEM samples and tomography samples are plate-like (Fig. 4d in a volume of $20 \times 10 \times 4$ μm$^3$). The plate-like morphology is similar to the $\beta\text{-}Al_5FeSi$ in Al-Si alloys reported in previous studies [4, 14]. Such a large faceted structure promotes the formation of porosity defects and act as the sites for crack initiation during mechanical loading [17].

In this paper, the true 3D network structures and morphologies of the Fe-rich intermetallic phases in Al-5.0Cu-0.6Mn alloys with Fe concentrations of 0.5% and 1.0% were studied and quantified for the first time. The Fe phases in the 1.0Fe alloy are complex 3D networks with well-developed branches of the node lengths of 5-25μm. While the Fe-phases in the 0.5Fe alloy have the similar 3D structures, but more compact and shorter branch node lengths (3-17 μm). The Fe phases in the 1.0Fe alloy also developed sharper curvatures than those in the 0.5Fe alloy with the standard deviation of the mean curvature distribution decreased from 2.71 to 1.33. Furthermore, the true 3D structures of the 4 different type Fe phases, $Al_3(FeMn)$, $Al_6(MnFe)$, $\alpha\text{-}Al_{15}(FeMn)_3Cu_2$, $\beta\text{-}Al_7Cu_2Fe$ in the two alloys are revealed and demonstrated.

**Acknowledgements**

Authors gratefully acknowledge the support from UK-EPSRC grants (EP/L019884/1, EP/L019825/1, EP/L019965/1), Natural Science Foundation of China (51374110), and Team project of Natural Science Foundation of Guangdong Province (2015A030312003). We acknowledge the Paul Scherrer Institut, Villigen, Switzerland for provision of synchrotron radiation beamtime at the TOMCAT beamline X02DA of the SLS under proposal number 20160284 and would like to thank Dr C. M. Schlepütz for assistance. We also would like to acknowledge the Viper High Performance Computing facility of the University of Hull and its support team. Financial support from the Chinese Scholarship Council (for Y. Zhao's PhD study at Hull University in Nov. 2016 - Nov. 2017) is also acknowledged.

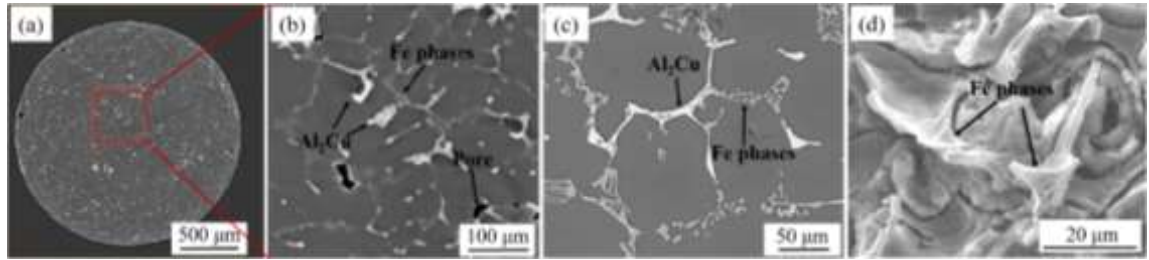

**Fig. 1.** (a) a typical 2D slice from the tomography scan of the 0.5Fe alloy; (b) the enlarged image of the framed area in (a) and processed using a 3D bilateral filter; (c) a typical SEM image of the 0.5Fe alloy; (d) a SEM image, showing the deeply-etched Fe phases.

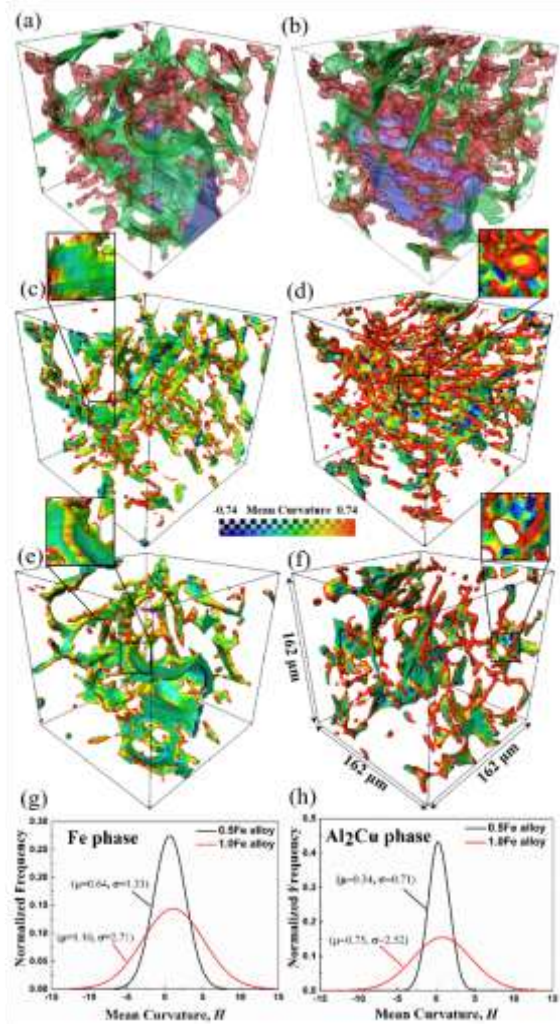

**Fig. 2** Typical 3D volume rendering of the Fe phases and Al$_2$Cu phases and their mean curvatures in (a) the 0.5Fe alloy and (b) the 1.0Fe alloy (red: Fe phases, green: Al$_2$Cu phases, blue: α-Al matrix). 3D morphology coloured with its mean curvature: (c) Fe phases in the 0.5Fe alloy; (d) Fe phases in the 1.0Fe alloy; (e) Al$_2$Cu phases in the 0.5Fe alloy; (f) Al$_2$Cu phases in the 1.0Fe alloy. The distributions of the mean curvatures of: (g) the Fe phases, and (h) the Al$_2$Cu phases, respectively.

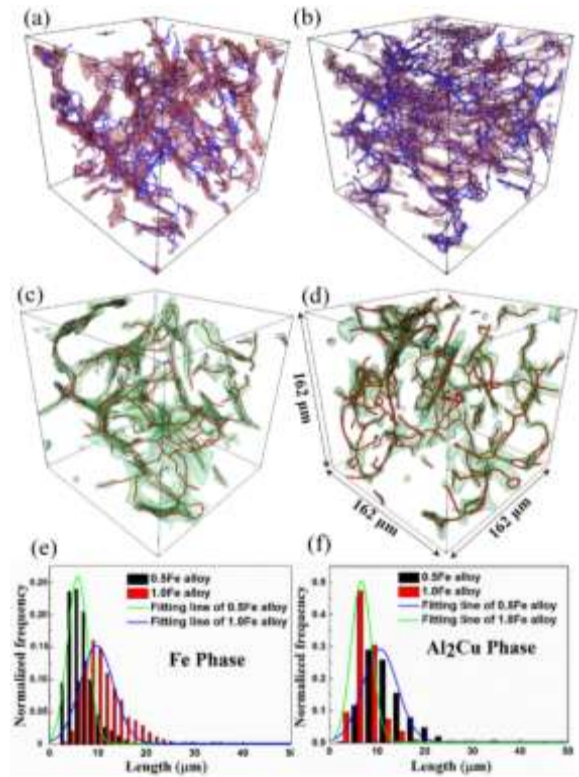

**Fig. 3** Skeletons of the Fe phases in: (a) the 0.5Fe alloy and (b) the 1.0Fe alloy; skeletons of the Al$_2$Cu phases in (c) the 0.5Fe alloy and (d) the 1.0Fe alloy; (e) and (f) the distributions of the node lengths of the Fe phases and Al$_2$Cu phases, respectively.

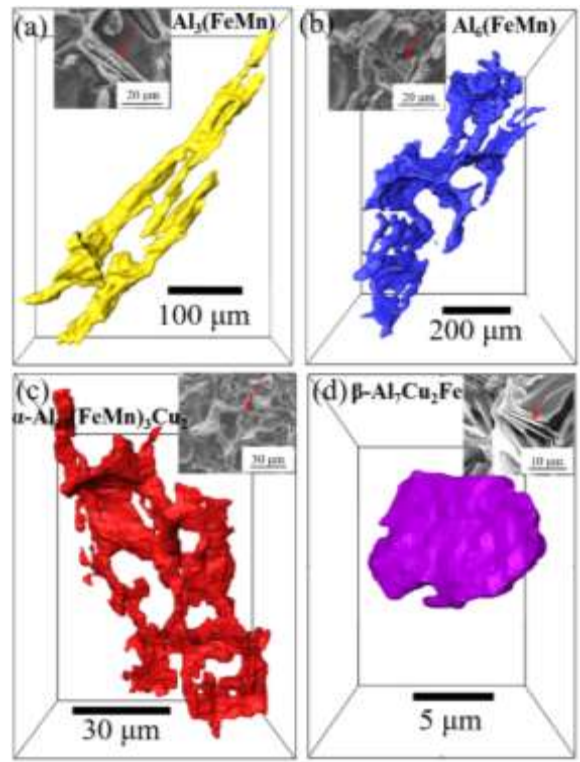

**Fig. 4.** 3D structures of 4 different Fe phases and their SEM images (insets) of deeply-etched morphologies: (a) rod-like $Al_3(MnFe)$; (b) Chinese script type $Al_6(MnFe)$; (c) Chinese script type $α-Al_{15}(FeMn)_3Cu_2$; (d) plate-like $β-Al_7Cu_2Fe$.